 \documentclass[pre,twocolumn,showpacs,preprintnumbers,amsmath,amssymb]{revtex4} 

\newcommand{\paper}[4]{#1 {\bf #2}, #3 (#4)}

\newcommand{\pr}[3]{\paper{Phys. Rev.}{#1}{#2}{#3}}
\renewcommand{\prl}[3]{\paper{Phys. Rev. Lett.}{#1}{#2}{#3}}
\renewcommand{\pra}[3]{\paper{Phys. Rev. A}{#1}{#2}{#3}}

\renewcommand{\pre}[3]{\paper{Phys. Rev. E}{#1}{#2}{#3}}

\newcommand{\sci}[3]{\paper{Science}{#1}{#2}{#3}}

\newcommand{\physrep}[3]{\paper{Phys. Rep.}{#1}{#2}{#3}}

\newcommand{\physletta}[3]{\paper{Phys. Lett. A}{#1}{#2}{#3}}

\newcommand{\epjd}[3]{\paper{Eur. Phys. J. D}{#1}{#2}{#3}}

\newcommand{\physicaa}[3]{\paper{Physica A}{#1}{#2}{#3}}

\newcommand{\applphya}[3]{\paper{Appl. Phys. A}{#1}{#2}{#3}}

\newcommand{\jphysoneF}[3]{\paper{J. Phys. I France}{#1}{#2}{#3}}

\newcommand{\jphysthreeF}[3]{\paper{J. Phys. III France}{#1}{#2}{#3}}

\renewcommand{\josab}[3]{\paper{J. Opt. Soc. Am. B}{#1}{#2}{#3}}

\newcommand{\ibid}[3]{\paper{\textit{ibid.}}{#1}{#2}{#3}}

\newcommand{\book}[5]{\textit{#1}, #2#3, #4 (#5)}

\input epsf
\usepackage{graphicx}
\usepackage{dcolumn}
\usepackage{bm}

\newcommand{\infig}[2]{\begin{center}
                                    \mbox{ \epsfxsize #1 \epsfbox{#2}}
                                      \vspace{-0.8cm}
                                    \end{center}}

\newcommand{\be}{\begin{equation}}
\newcommand{\ee}{\end{equation}}
\newcommand{\bea}{\begin{eqnarray}}
\newcommand{\eea}{\end{eqnarray}}
\newcommand{\eqt}[1]{\mbox{$#1$}}

\renewcommand{\mathbf}[1]{\mbox{\boldmath $#1$}}

\newcommand{\ie}{{\it i.e.}~}
\newcommand{\eg}{{\it e.g.}~}

\newcommand{\bfig}{\begin{figure}[!ht]}
 \newcommand{\efig}{\end{figure}}

\newcommand{\bbib}{\begin{thebibliography}{99}}
 \newcommand{\ebib}{\end{thebibliography}}

\begin{document}

\title{Directed transport of Brownian particles in a double symmetric potential}

\author{Laurent Sanchez-Palencia\footnote{Present adress:Institut f\"ur Theoretische Physik, Universit\"at Hannover, D-30167 Hannover,Germany}
}
\affiliation{Laboratoire Kastler Brossel, D\'epartement de Physique de l'Ecole
Normale Sup\'erieure, 24, rue Lhomond, 75231, Paris cedex 05,
France.}

\date{\today}

\begin{abstract}
We investigate the dynamics of Brownian particles in internal state- dependent symmetric and periodic potentials. Although no space or time symmetry of the Hamiltonian is broken, we show that directed transport can appear. We demonstrate that the directed motion is induced by breaking the symmetry of the transition rates between the potentials when these are spatially shifted. Finally, we discuss the possibility of realizing our model in a system of cold particles trapped in optical lattices.
\end{abstract}


\maketitle


In usual Brownian motion \cite{brownian}, thermal fluctuations act as a reservoir of energy which distillation within the system of particles results in random trajectories and molecular chaos on a microscopic scale and in symmetric spatial diffusion on a macroscopic scale. The problem of converting the energy of noise into deterministic mechanical work (\ie {\it rectification} of the energy of noise) to induce directed motion was first addressed by Schmoluchowski and Brillouin \cite{rectification} and has attracted much attention during the past years (see Ref.~\cite{motors02} and references therein). This question is crucial in particular for understanding how molecular pumps work in biological environnements \cite{biology} or for designing new synthetic pumps for chemistry \cite{chemistry}. It is by now clear \cite{reimann02} that the necessary and generally sufficient conditions for noise rectification are : (i) breaking the spatio-temporal symmetry of the Hamiltonian (to match the Curie principle \cite{curie94}) and (ii) non thermodynamic equilibrium situations (to avoid detailled balance and resulting zero current \cite{prost94}).

Most studies rely on Brownian particles experiencing a spatially periodic potential \eqt{u (x)} plus eventually a time \eqt{\tau}- periodic force \eqt{f (t)}. After Curie's principle \cite{curie94}, if both symmetry conditions
\bea
& & \eqt{u (x) = u (2 x_0 - x)} \label{eq:spacesymm} \\
& & \eqt{f (t+\tau/2) = -f (t)} \label{eq:timesymm}
\eea
are fulfilled (\eqt{x_0} being any given position), then no directed motion is possible. On the contrary, breaking one of these symmetries is generally sufficient to ensure a net motion of the particles \cite{reimann02}. Both cases of spatial symmetry~(\ref{eq:spacesymm}) breaking \cite{prost94,spaceratchet} and temporal symmetry (\ref{eq:timesymm}) breaking \cite{timeratchet} have been widely studied in a large variety of systems and evidences for directed transport have been given. In these systems, the non-equilibrium requirement is ensured by either non-gaussian noise, temperature oscillations or flashing potentials \cite{reimann02}. In recent years, more elaborated systems that mimic biological system more properly have been investigated. It has been thus demonstrated that interestingly directed motion can be obtained in symmetric potentials and in the absence of a time-dependant force, \ie in systems where conditions (\ref{eq:spacesymm})-(\ref{eq:timesymm}) are fulfilled. This requires complex situations, \eg position-dependent mobility in flashing potentials \cite{luchsinger00}, state-dependent mobility \cite{dan01} or time-dependant spatial fluctations of walls \cite{litchenberg04}.

In this work, we propose a very simple system for which directed motion is obtained despite validity of both spatial~(\ref{eq:spacesymm}) and temporal~(\ref{eq:timesymm}) conditions. Here, it is the whole dynamics made of Hamiltonian motion phases and dissipative processes which is non-symmetric. More precisely, directed motion is induced for low-damped particles in internal state-dependent periodic potentials with spatially shifted pinning points and for different transition rates between the potential curves. The behavior is explained on symmetry grounds and with the help of a simple rectification process that qualitatively describes the dynamics of the particles. Our conclusions are supported by a study of the influence of the relevant parameters (shift between the potentials and strengths of the transtion rates). We finally discuss the possibility of experimental observation of such an effect in a system of cold particles trapped in optical lattices.


\bfig
\infig{25em}{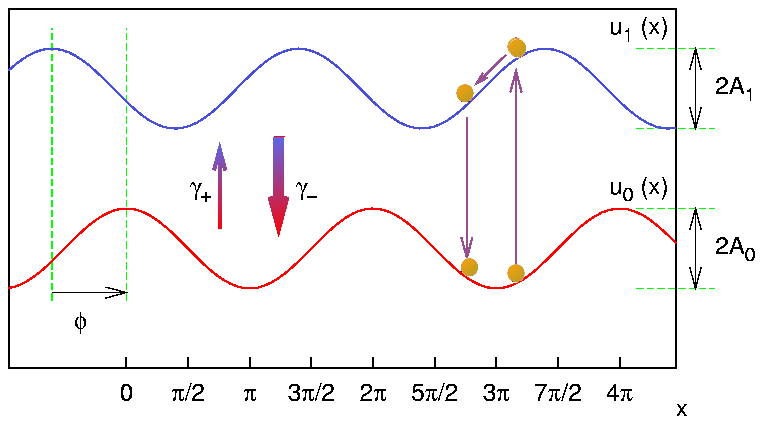}
\caption{Diagram of the system considered in this work. A Brownian particle can live in two different internal states \eqt{0} and \eqt{1} corresponding to \eqt{\phi}-shifted potentials; the transition rates are asymmetric (\eqt{\gamma_+ \neq \gamma_-}). The rectification process is shown in the right-hand side of the figure (see text below).}
\label{fig:system}
\efig

Consider a Brownian particle living in \eqt{N} internal states (indexed by \eqt{j \in [0 ..N-1]}) corresponding (eventually) to different momentum diffusion constants \eqt{D_{\textrm{v} j} (x)} and different spatially modulated potentials \eqt{u_j (x)} and submitted to a time-dependent homogeneous force \eqt{f_j (t)}. For example, the different potentials may correspond to different chemical species and the transitions to chemical reactions or to different quantum internal atomic states and light-induced transitions respectively.

The Fokker-Planck equation (FPE) for the population \eqt{W_j (x,v,t)} of the state characterized by \eqt{x,v} (external state) and \eqt{j} (internal state), reads \cite{risken89}:
\bea
& & \left[ \partial_t + v \partial_x \right] W_j + \partial_v \left[ v + u'_j (x) - f_j (t) + D_{\textrm{v} j} (x) \partial_v \right] W_j \nonumber \\
& & = \sum_{j' \neq j} \gamma_{j' \rightarrow j} (x)
W_{j'} - \sum_{j' \neq j} \gamma_{j \rightarrow j'} (x) W_{j}
\label{eq:fpe}
\eea
where \eqt{\gamma_{j' \rightarrow j}} represents the transition rate from internal state \eqt{j'} to state \eqt{j}. The FPE~(\ref{eq:fpe}) is written in time units of the friction coefficient and in space units of the typical variation scale of the potentials \eqt{u_j} so that all variables are dimensionless. The left-hand side of equation~(\ref{eq:fpe}) is a usual FPE for a particle submitted to a friction force, a potential-induced force, a time-dependent force and a momentum diffusion and thus captures the one dimensional Brownian motion in a regime ranging from overdamping to low damping. The left-hand side of equation~(\ref{eq:fpe}) is a source term that accounts for random jumps between the internal states. Non-thermal transition rates (\eg induced by chemical reactions or optical transitions) result in non-thermodynamic equilibrium situations \cite{prost94}. In the following, we suppose that
\begin{enumerate}
\item[-] \eqt{u_j} is spatially periodic and symmetric in the sense of~(\ref{eq:spacesymm}),
\item[-] \eqt{f_j} is \eqt{\tau}- periodic and fulfills~(\ref{eq:timesymm}); note that the average value of \eqt{f_j} vanishes: \eqt{\frac{1}{\tau} \int_0^\tau \textrm{d}t f_j (t) = 0}.
\end{enumerate}

We consider the case of a double symmetric potential (\eqt{N=2}) with (\eqt{\phi}-) shifted maxima (see Fig.~\ref{fig:system}) with (in general) non-symmetric transition rates (\eqt{\gamma_{0,1} \neq \gamma_{1,0}}) and in the absence of a time-dependent force (\eqt{f_j \equiv 0}). More precisely, we use sine potentials with the same period \eqt{2 \pi}:
\bea
u_0 (x) & = & A_0 \cos (x) \nonumber \\
u_1 (x) & = & A_1 \cos (x + \phi) ~.
\label{eq:potentials}
\eea
We assume that the mean kinetic energy (proportionnal to the temperature) of the system is smaller than the depth of the trapping wells corresponding to the potential minima (\ie typically \eqt{D_\textrm{v} < A_j}). Moreover, we assume the energy damping to be much smaller than the typical frequency of the oscillations in the potential wells (\ie \eqt{\sqrt{|A_j|} \gg 1}) so that the dynamics falls into the low damping regime. Thus the particles are essentially trapped in the wells and their basic behavior consists in weakly damped and noisy oscillations at the bottom of the potential minima.

For the sake of simplicity, we finally assume minimal requirements for the phenomenon of directed transport demonstrated in the reminder of the paper:
\begin{enumerate}
\item[-] \eqt{A_0 = A_1 = A},
\item[-] \eqt{D_{\textrm{v} j}} is position- and internal state- independent (\eqt{D_{\textrm{v} j} = D_\textrm{v}}),
\item[-] \eqt{\gamma_+}(\eqt{=\gamma_{0 \rightarrow 1}}) and \eqt{\gamma_-}(\eqt{=\gamma_{1 \rightarrow 0}}) are position
         independent and different in general.
\end{enumerate}

Focus first on the Hamiltonian part of the dynamics. It is clear that the Hamiltonian is symmetric with respect to both space and time. Indeed, it is invariant through \eqt{\left\{ u_\epsilon (x) \right\} \leftrightarrow \left\{ u_{1-\epsilon} (2 x_0 -x) \right\}}
where \eqt{x_0} is the middle point between a maximum of \eqt{u_0} and a maximum of \eqt{u_1}. Moreover, there is no explicit time dependence in the equations of motion~(\ref{eq:fpe}). Therefore, the Hamiltonian part of the dynamics is clearly symmetric with respect to both conditions~(\ref{eq:spacesymm}) and (\ref{eq:timesymm}) and cannot induce any directed motion.
However, the dynamics has to be regarded as an intercombination of Hamiltonian motion (motion on the potential curves) and dissipative processes (transitions between different internal states). Here, the system no longer preserves the spatio-temporal symmetry since it is not invariant through \eqt{\left\{ u_\epsilon (x), \gamma_\epsilon \right\} \leftrightarrow \left\{ u_{1-\epsilon} (2 x_0 -x), \gamma_{1-\epsilon} \right\}} except for \eqt{\gamma_+ = \gamma_-}. Hence different transition rates break the spatio-temporal symmetry of the system and force-free directed motion can be expected. It is the aim of the following paragraphs (i) to demonstrate this phenomenon and (ii) to describe the corresponding microscopic mechanism.


We performed numerical simulations of the dynamics of the particles in the system described above by means of a Monte-Carlo integration with a set of typically \eqt{2000} atoms. It is then straightforward to determine the mean position of the cloud of particles as a function of time from which we deduce the average velocity (\eqt{V}) of the cloud:
\be
V = \frac{\textrm{d}\langle x \rangle}{\textrm{d} t}
\label{eq:dynamics}
\ee
where \eqt{\langle~.~\rangle} stands for the ensemble average over all particles.
For all the results presented in this work, we found a constant mean velocity (see inset of Fig.~\ref{fig:phi}).

We first study the behavior of the mean velocity \eqt{V} as a function of the space shift \eqt{\phi} between the potentials \eqt{u_0} and \eqt{u_1} in a situation where the transition rates between the internal states are significantly different (\eqt{\gamma_- = 3 \gamma_+}). The results of the numerical simulations are shown in Fig.~\ref{fig:phi}. The cloud of particles clearly shows a directed motion through the potentials for almost all values of \eqt{\phi}. These results show that a directed transport of low-damped Brownian particles can be induced in a $2$ state spatially symmetric potentials with shifted minima. The main features of Fig.~\ref{fig:phi} can be explained under symmetry considerations. First, the average velocity vanishes (\eqt{V = 0}) only for discrete values of \eqt{\phi} corresponding to symmetric situations (\eqt{\phi = 0, \pi}). Here the symmetry is different from the one discussed above and reads \eqt{\left\{ u_\epsilon (x), \gamma_\epsilon \right\} \leftrightarrow \left\{ u_\epsilon (-x), \gamma_\epsilon \right\}}. Note that this space inversion is valid only for \eqt{\phi = 0} or \eqt{\phi = \pi} and thus \eqt{V = 0} only for these values of \eqt{\phi}. Second, changing \eqt{\phi} into \eqt{(2\pi)-\phi}, the shift of the minima of \eqt{u_1} with respect to the minima of \eqt{u_0} gets opposed and this results in the inversion of the direction of motion of the particles as observed on Fig.~\ref{fig:phi}. Third, we find that the mean speed (\eqt{|V|}) of the directed transport is maximum for values of \eqt{\phi} close to \eqt{\pi/2} or \eqt{3\pi/2}, \ie in a situation where the minima of the potentials are almost maximally shifted.

\bfig
\infig{25em}{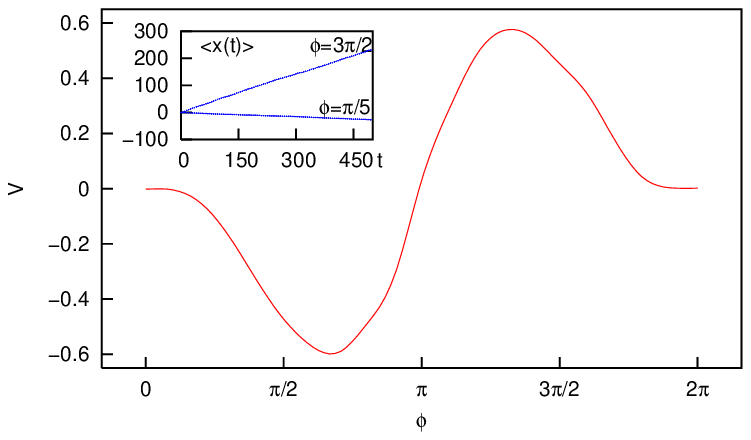}
\caption{Mean velocity of the particles as a function of the phase \eqt{\phi}.
Inset: displacement of the particles as a function of time for two different values of \eqt{\phi}.
Numerical parameters for this simulation are \eqt{A=200}, \eqt{D_\textrm{v}=75}, \eqt{\gamma_- = 0.75 \Gamma}, \eqt{\gamma_+ = 0.25 \Gamma} and
\eqt{\Gamma = 10}. }
\label{fig:phi}
\efig

We now describe the noise rectification mechanism. In the low damping regime considered in this work, the microscopic dynamics of the particles consists in noisy oscillations at the bottom of the potential wells interrupted by random transitions from a potential curve to another. Suppose now that \eqt{0 < \gamma_+ < \gamma_-}. Then a particle spends the most of time in potential \eqt{u_0} and oscillates around a given minima. Due to the finite value of \eqt{\gamma_+}, it may be transfered into potential \eqt{u_1}. If the minima and maxima of \eqt{u_0} and \eqt{u_1} are spatially shifted (\eqt{\phi \neq n \pi}, with \eqt{n} an integer), the particle initially localized near a minimum of \eqt{u_0} falls away from a minimum of \eqt{u_1} and consequently experiences a force oriented towards the left (for \eqt{2n \pi < \phi < (2n+1) \pi}) or towards the right (for \eqt{(2n-1) \pi < \phi < 2n \pi}). It thus moves in the corresponding direction. It is finally rapidly pumped back to potential \eqt{u_0}, generally in the initial trapping well. As a consequence of this cycle (see right hand side of Fig.~\ref{fig:system}), a mean directed force is exerted on the particle and the center of the oscillations of this particle in a well of \eqt{u_0} is shifted (towards the left for \eqt{2n \pi < \phi < (2n+1) \pi} or towards the right for \eqt{(2n-1) \pi < \phi < 2n \pi}). This formally lowers the potential barrier for escaping the trapping well in the corresponding direction and rises it in the opposite direction. This clearly induces a directed average motion of the cloud of particles. Note that this mechanism matches the mean features of Fig.~\ref{fig:phi} described above.

As discussed previously, considering different values of the transition rates \eqt{\gamma_+ \neq \gamma_-} is crucial to break the spatio-temporal symmetry of the system and correspondingly to induce directed motion of the particles through the periodic structure. We now investigate the influence of the transition rates on the average velocity \eqt{V}. We plot \eqt{V} as a function of \eqt{\gamma_+} on Fig.~\ref{fig:gamma} for three different values of the total transition rate \eqt{\Gamma = \gamma_+ + \gamma_-} and in a situation where the potentials \eqt{u_0} and \eqt{u_1} are significantly shifted (\eqt{\phi = 3 \pi / 2}). We find that \eqt{V} strongly depends on \eqt{\gamma_+ / \Gamma} as expected. For \eqt{\gamma_+/\Gamma = 0} (\eqt{\gamma_-/\Gamma = 1}) or \eqt{\gamma_+/\Gamma = 1} (\eqt{\gamma_-/\Gamma = 0}), the system is equivalent to a Brownian particle in a single symmetric potential and therefore does not show any directed motion: \eqt{V = 0} \cite{curie94}. \eqt{\gamma_+ / \Gamma =0.5} (\eqt{\gamma_+ = \gamma_-}) corresponds to a symmetric situation as discussed above and results in no directed motion as confirmed by Fig.~\ref{fig:gamma}. For all other values of \eqt{\gamma_\pm / \Gamma}, a directed motion is clearly demonstrated. The maximum of \eqt{|V|} is found to correspond to strongly asymmetric values of \eqt{\gamma_+} and \eqt{\gamma_-} (typically for \eqt{\gamma_\pm / \gamma_\mp \simeq 8}). This is consistent with the rectification mechanism proposed above. It is indeed clearly necessary that particles get trapped in the wells of \eqt{u_{\epsilon}} (\eqt{\epsilon = 0, 1}) and encounter short transitions in \eqt{u_{1-\epsilon}} to induce significant directed motion. We finally note that \eqt{\gamma_+/\Gamma \leftrightarrow 1-\gamma_+/\Gamma} results in \eqt{V \leftrightarrow -V}. This is an obvious consequence of the anti-symmetry of the system through exchange of the internal states \eqt{0 \leftrightarrow 1} and corresponding change of the phase shift: \eqt{\phi \leftrightarrow -\phi}. We finally note that \eqt{|V|} decreases when \eqt{\Gamma} increases and this indicates that the rectification process is no longer efficient for high values of \eqt{\Gamma}. Indeed, in this case, a particle undergoes many cycles so that it experiences an average potential \eqt{(\gamma_- u_0 + \gamma_+ u_1)/\Gamma} which is symmetric and again no directed motion is possible \cite{curie94}.

\bfig
\infig{25em}{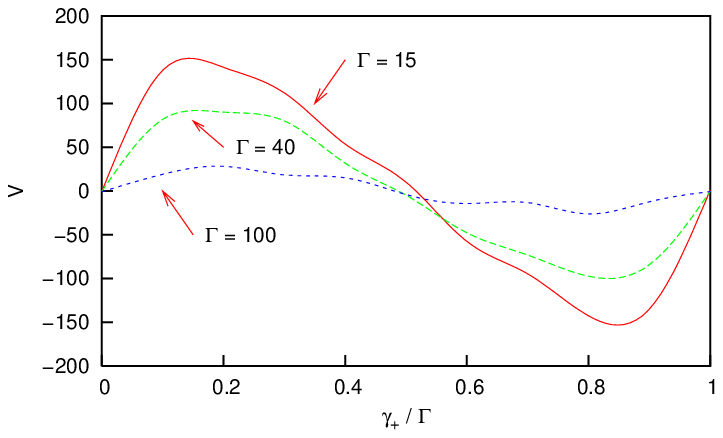}
\caption{Average velocity \eqt{V} of the cloud of particles as a function of the transition rates between internal states. \eqt{V} is plotted versus \eqt{\gamma_+} for a fixed value of \eqt{\Gamma = \gamma_+ + \gamma_-}.
Parameters of the simulation are \eqt{A=200}, \eqt{D_\textrm{v}=75} and \eqt{\phi = 3 \pi /2}.}
\label{fig:gamma}
\efig

We now discuss the possibility of realizing our model in experimental setups. Ratchet effects have been studied in a large variety of systems \cite{motors02} but few of them allow for independant control of the physical parameters in wide ranges with accuracies as good as in systems of cold atoms trapped in optical lattices \cite{robilliard_ratchets,schiavoni03}. We thus propose to use such a system to observe the phenomenon of directed transport in a double symmetric potential. Optical lattices are formed by the interference pattern of laser beams \cite{grynberg01}. In appropriate configurations, matter-light interactions create (i) a periodic and symmetric potential for the atoms, (ii) a friction-like force and (iii) random fluctuating forces. The competition between the last two phenomena results in an equilibrium kinetic energy lower than the potential depth. The atomic dynamics has been shown to be equivalent to Brownian motion in a periodic potential \cite{sisyphus89,sanchez02_D} in the low-damping regime \cite{sanchez02_SR}. Therefore these properties match the requirements considered in this work for observing directed transport in a double symmetric potential. In fact, the parameters used for the simulations presented in this work roughly mimic optical lattices. In order to induce a double potential, we suggest to follow the scheme recently presented in Ref~\cite{doublelattice}. By illuminating $^{133}$Cs with two lasers respectivelly near-resonant with two atomic transitions, the atoms see a double sine-like potential. Moreover, the atom-light interaction induces non-thermal transition rates between these potentials. It is finally worth noticing that the space shift between the potentials, the potential depths and the transition rates can be controlled experimentally. Therefore, we believe that this system matches the main characteristics of the model studied in this work (phase-shifted symmetric potentials, non-thermal and different transition rates, temperature below the potential depth, low-damping regime) and should be well suited for observing directed motion.

To sum up, we have studied the directed motion of particles induced by spatio-temporal symmetry breaking in a multi-level periodic system. Our system consists in low-damped Brownian particles in a multi-level flashing periodic and symmetric potential. We have shown that directed motion is induced whenever the minima of the different potentials are shifted and for different transition rates. This breaks the spatio-temporal symmetry of the complete dynamics of the particles which corresponds to a combination of Hamiltonian motion (on the potential curves) and dissipation (related to the internal state transitions). We have finally proposed a system based on cold atoms trapped in an optical lattice where directed transport in a double symmetric potential may be experimentally observed and characterized with a high degree of accuracy.

Exploring Brownian motors and corresponding ratchet effect helps in understanding the physics of protein motors, Josephson junction arrays and quantum dots. Usual models consider particles in space asymmetric potentials \cite{spaceratchet} or undegoing time-asymmetric forces \cite{timeratchet}. More recent models consider elaborated schemes in symmetric potentials \cite{luchsinger00,dan01,litchenberg04}. In this work, we have proposed a very simple scheme in which ratchet effect can be observed in symmetric potentials. In fact our system considers a low-damped Brownian particle minimally in a 2-state potential with shifted minima and different transition rates. This simplicity makes the scheme quite general and it could be observed in various systems for example in cold atoms optical lattices.

We are indebted to Professor Anders Kastberg for enlightening discussions on the physics of double optical latices and for valuable comments on the paper.
Laboratoire Kastler Brossel is an ``unit\'e mixte de recherche de l'Ecole Normale Sup\'erieure et de l'Universit\'e Pierre et Marie Curie associ\'ee au Centre National de la Recherche Scientifique (CNRS)''.


\end{document}